\documentclass[preprints,article,accept,moreauthors,pdftex]{mdpi}
\firstpage{1} 
\pubvolume{1}
\issuenum{1}
\articlenumber{0}
\pubyear{2023}
\copyrightyear{2023}
\datereceived{27.10.2023 } 
\daterevised{29.10.2023 } 
\dateaccepted{ } 
\datepublished{ } 
\hreflink{https://doi.org/} 
\pdfoutput=1 

\Title{A BRAIN study to tackle image analysis with artificial intelligence in the ALMA 2030 era}

\TitleCitation{A BRAIN study to tackle image analysis}

\Author{Fabrizia Guglielmetti $^{1}$, Michele Delli Veneri $^{2}$, Ivano Baronchelli $^{3,4}$, Carmen Blanco$^{5}$, Andrea Dosi $^{6}$, Torsten En\ss lin $^{5}$, Vishal Johnson $^{5}$, Giuseppe Longo $^{6}$, Jakob Roth$^{5,7}$, Felix Stoehr $^{1}$, \L ukasz Tychoniec $^{1}$, Eric Villard $^{1}$}

\longauthorlist{yes}

\AuthorNames{Fabrizia Guglielmetti, Michele Delli Veneri, Ivano Baronchelli, Carmen Blanco, Andrea Dosi, Torsten En\ss lin, Vishal Johnson, Giuseppe Longo, Jakob Roth, Felix Stoehr, \L ukasz Tychoniec and Eric Villard}

\AuthorCitation{Guglielmetti F. et al.}

\address{%
$^{1}$ \quad European Southern Observatory, Karl-Schwarzschild-Straße 2, D-85748 Garching bei M\"unchen \\
$^{2}$ \quad INFN Section of Naples (INFN), Via Cinthia, I-80126, Naples \\
$^{3}$ \quad INAF - Institute of Radio Astronomy, Via Gobetti 101, I-40129 Bologna\\
$^{4}$ \quad Italian ALMA Regional Centre, Via Gobetti 101, I-40129 Bologna\\
$^{5}$ \quad Max Planck Institute for Astrophysics, Karl-Schwarzschild-Straße 1, D-85748 Garching bei M\"unchen\\
$^{6}$ \quad Federico II University, Via Cinthia, I-80126 Fuorigrotta, Napoli\\
$^{7}$ \quad Technical University of Munich, Boltzmannstr. 3, D-85748 Garching bei M\"unchen}
\corres{Correspondence: gugliel@eso.org}

\abstract{An ESO internal ALMA development study, BRAIN, is addressing the ill-posed inverse problem of synthesis image analysis employing astrostatistics and astroinformatics. These emerging fields of research offer interdisciplinary approaches at the
intersection of observational astronomy, statistics, algorithm development, and data science. In this study, we provide evidence of the benefits of employing these approaches to ALMA imaging for operational and scientific
purposes. We show the potential of two techniques, RESOLVE and DeepFocus, applied to ALMA calibrated science data. Significant advantages are provided with the prospect to improve the quality and completeness of the data products
stored in the science archive and overall processing time for operations. 
Both approaches evidence the logical pathway to address the incoming revolution in data rates dictated by the planned electronic upgrades. Moreover, we bring to the community additional products through a new package, ALMASim, to promote advancements in these fields, 
providing a refined ALMA simulator usable by a large community for training and/or testing new algorithms.
}


\usepackage{makeidx}
\makeindex 
\begin{document}

%


\section{Introduction}

\begin{figure}
\vspace{-4cm}
\centering
\rotatebox{-90}{\includegraphics[width=10.cm]{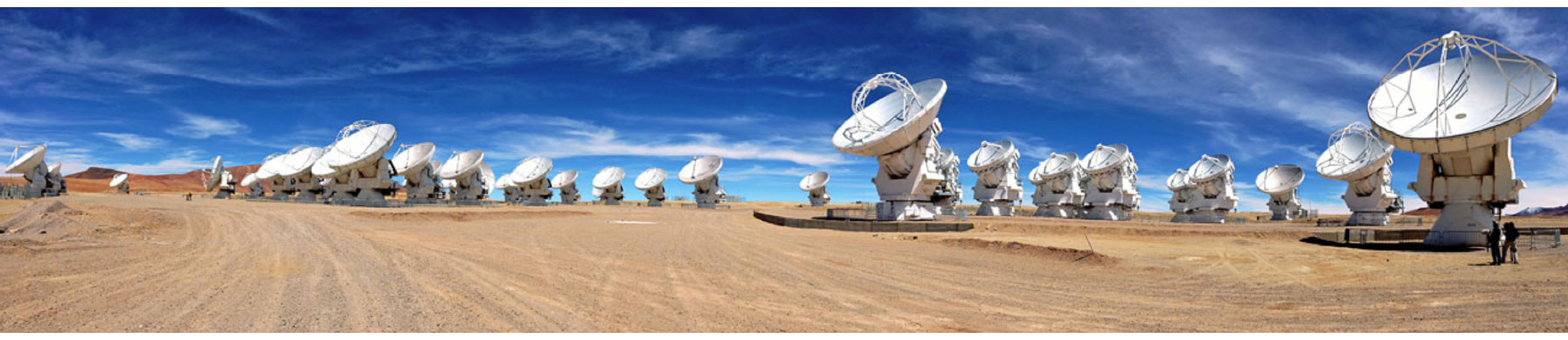}}
\vspace{-3cm}
\caption{ALMA antennas on the Chajnantor plateau. Credit: ESO.}
\label{almaimg}
\end{figure}

Since its inception in 2011, the Atacama Large Millimeter/submillimeter  Array (ALMA) (Fig.~\ref{almaimg}), see e.g.~\cite{Wootten2009}, has been at the forefront of astronomical research, consistently delivering groundbreaking scientific discoveries with its versatile observatory capabilities. In 2018, the ALMA Development Roadmap was providing a forward-thinking perspective to increase the observatory's capabilities out to 2030 (ALMA2030) \cite{Carpenter2018,Carpenter2020}. 
Scientifically motivated by fostering the understanding of the origin of galaxies, 
the origin of chemical complexity 
and the origins of planets, ALMA is going to be enhanced addressing  (1) larger bandwidths and better receiver sensitivity, (2) improvements to the ALMA Science Archive, (3) longer baselines. 
In addition, the observatory is currently expanding the frequency coverage to (35-950) GHz with two new receivers (band 1 and band 2)  to capture a broader range of celestial objects and phenomena.

The ALMA2030 Roadmap \cite{Carpenter2018} encourages innovative technical concepts rooted in astronomy, even if they are not explicitly listed within the document. In this sense, the ESO internal ALMA development study BRAIN fits into the original strategic vision of ALMA2030. In this study, synthesis imaging algorithms capable to learn from ALMA data are developed and/or tested. The powerful attribute of modern artificial intelligence systems enables algorithms to make data-driven decisions, discover insights, and provide predictions. We include the exploration of algorithms that can leverage Graphics Processing Units (GPUs) to maximize computational power, achieving high performance. GPUs  allow for more efficient and cost-effective computing solutions and opens up new possibilities for solving complex problems on large data volume. Moreover, we provide to the community a data base and software to create ALMA simulated data to promote the development of new synthesis imaging algorithms. 

\subsection{Data rate and data volume increment with ALMA2030}
The currently top priority of the ALMA2030 vision is to broaden the receiver Intermediate Frequency (IF) bandwidth by a factor of 2--4 (currently is $\sim$8 GHz) while upgrading the associated electronics and the correlator. A wider bandwidth will allow the observatory to capture more data in a shorter amount of time, accelerating data collection and analysis. Upgrades to the receivers and electronics improve the sensitivity, with the aim to detect fainter signals and gather more precise data. Efforts are made to reduce system noise, interference, and instrumental effects in order to refine the signal-to-noise ratio.
These enhancements have been officially designated as the ALMA2030 Wideband Sensitivity Upgrade  (WSU) \cite{Carpenter2022}. 

The development of the ALMA second-generation correlator, the improved receivers, wider bandwidth and transmission system will enable a substantial increase of the data volume during a single observation and up to 10 times more observing speed efficiency. When ALMA collected 1.7 PB of visibility data and products until Cycle 9, it is expected that ALMA2030 might augment the visibility data up to a factor of 100. 
Secondly, the ALMA Science Archive requires improvements in order to handle the data from the increased bandwidth in addition to the enhancement of a user-friendliness of the Archive and data products. 
Thirdly, the maximum baselines may be increased by a factor of 2–3 and the number of antennas will grow up to 16\% . Consequently, ALMA resolution and uv coverage will enrich from today' several possible solutions because of longest baselines and more efficient observational starategies. 

The expanded bandwidth directly leads to enhancements in both continuum sensitivity and spectral coverage. 
While the exact capabilities that can be offered are not yet known, an increase of the data-rate between $10-100$ times is envisaged. 
Due to the increased sensitivity and data volume, continuum identification can become an awkward procedure especially for the potential increment of spectral line detection. 
During the ramp up years of the WSU, observations will occur with a lower number of antennas than employed today. In this time frame, sparse sampling may affect the data.   
To handle the increased volume and/or complexity of the data, data processing and analysis tools have to be strategically developed. 

\section{Methods and Results: Artificial Intelligence for synthesis imaging with BRAIN}
 The ESO internal ALMA  development study named BRAIN has been proposed in late 2019 and will get to conclusion by the end of 2023. 
BRAIN stands for Bayesian Reconstruction through Adaptive Image Notion. Alternatives to commonly used applications in image processing are sought and tested \cite{Guglielmetti2019,Guglielmetti2022}. 
This study explores astrostatistics and astroinformatics approaches suited to ALMA2030 era. 
Astrostatistics and astroinformatics are essential for advancing our understanding of the universe and for making sense of the enormous volumes of data collected by astronomical instruments and observatories \cite{Aneta2019}. 

Astrostatistics involves the application of statistical analysis, data modeling, and probability theory to solve problems, performing predictions and providing robust uncertainty quantification. 
Extracting meaningful insights from astronomical data, astrostatistics enhances the quality and reliability of astronomical research by providing quantitative and data-driven approaches.

Astroinformatics focuses on the management, analysis, and interpretation of large and complex astronomical datasets, combining elements of astronomy, computer science, and information technology. The field plays a crucial role in executing operations of large observatories in real time. It can support operations for sanity checks on the health of the observatory, as well as real time image processing.    

In this study, the information field theory \cite{IFT} approach RESOLVE \cite{Henrik2016, Tychoniec2022, Roth2023} and the astroinformatics technique DeepFocus \cite{DelliVeneri2023} are applied to ALMA calibrated data. A companion package of DeepFocus, named ALMASim, have been developed with the additional scope to provide simulated data as well as open software for the scientific community. ALMASim enables training and testing supervised machine learning (ML) algorithms, but also for reliability and quality assessments and comparisons of several techniques. 

\subsection{Resolve}
\begin{figure}[H]
\centering
\rotatebox{-90}{\includegraphics[width=9.cm]{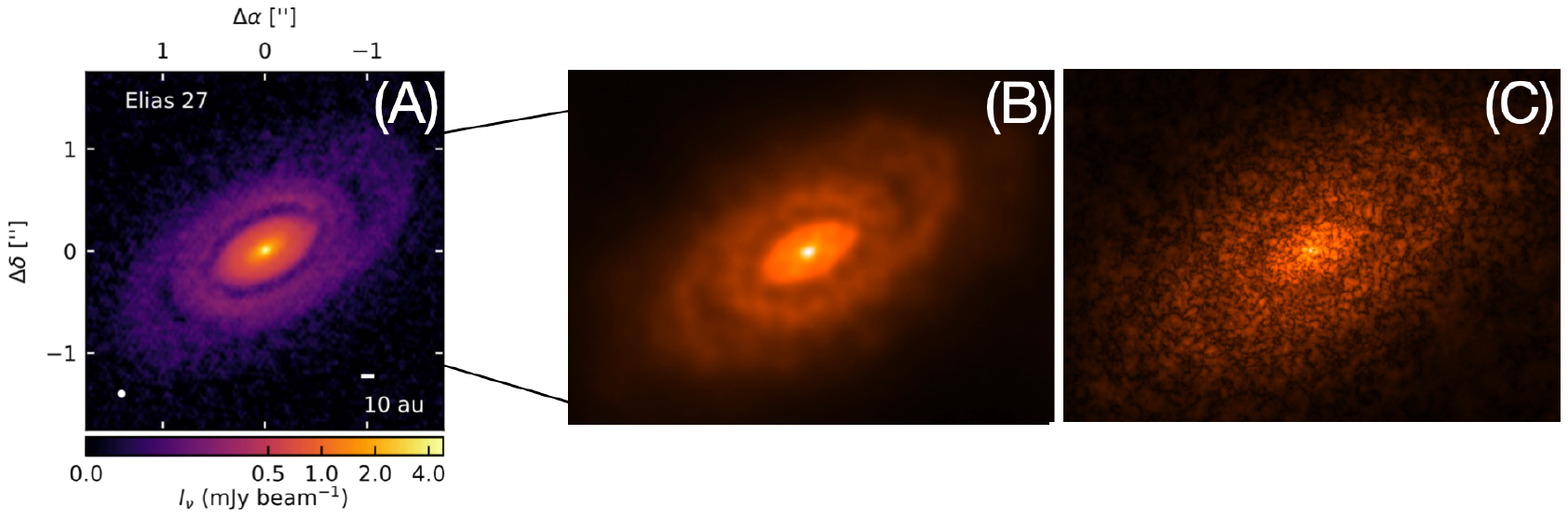}}
\vspace{-2cm}
\caption{Application of RESOLVE to Elias 27 from the DSHARP ALMA project at 240 GHz (1.25 mm) continuum. (A): The fiducial image as given by the DSHARP team \cite{Huang2018}, (B): RESOLVE mean sky map of Elias 27, (C) RESOLVE uncertainty map representation.}
\label{dsharp}
\end{figure}

The algorithm works on ALMA calibrated measurement sets in the uv space. The input data $d$ are modelled as a combination of log-celestial signal $s$ corrupted by the dirty beam $R$  (i.e.~the spatial response pattern of the telescope array during an observation) and the noise $n$ (systematic and random errors): $d=Re^{s}+n$. 
The process of synthesizing an image involves estimating the posterior probability density function (pdf) of potential true sky signal configurations using variational inference: $P(s|d)=\frac{e^{-H(s,d)}}{Z(d)}$. For more details on the RESOLVE algorithm see \cite{Henrik2016, Tychoniec2022, Roth2023} and references therein. The RESOLVE algorithm outputs posterior samples of sky images and power spectra. From these posterior samples, summary statistics such as a  mean sky map or an uncertainty map can be computed. 
The first application of RESOLVE to ALMA data is shown in \cite{Tychoniec2022}. In Fig.~\ref{dsharp}, an application of RESOLVE on a protoplanetary disk sample, Elias 27, from the ALMA Disk Substructures at High Angular Resolution Project (DSHARP) \cite{Andrews2018, Huang2018} is shown. In Fig.~\ref{dsharp}, image (A) is the fiducial image of the continuum detection of Elias 27 from the DSHARP data release \cite{DRel, Huang2018}: self-calibrated image employing CASA \cite{CASA}. Images (B) and (C) are the outcome of RESOLVE application to Elias 27 continuum ALMA data: the mean sky map and the uncertainty map, respectively. The DSHARP project aimed at detecting a large number of protoplanetary disks with high resolution (35 mas) to reveal amplitudes of small-scale substructures in the distributions of the disk material and understand their relation to planet forming process.
\vspace{-2cm}
RESOLVE is currently applied to other science cases, as self-calibration, Full Polarization, single dish imaging, as well as VLBI based filming of evolving black hole environments  \cite{arras2022variable, knollmuller2023resolving}. Those applications should be extended to ALMA data.

\subsection{DeepFocus}
The Deep Learning pipeline DeepFocus has been described in \cite{Guglielmetti2022, DelliVeneri2023}. It performs deconvolution, source detection and characterization. From the original development targeted to detect faint compact objects, the deconvolution algorithm has been improved to detect extended emissions with a new ML model, a meta-learner, capable to explore different architectures (e.g. CAE-VAE, U-Net, Res-Net) and assist in model selection to predict the most performing architecture given a task and a set of Interferometric data. A Bayesian optimization algorithm has been used for model selection. The procedure is supported by taxonomy. 
This framework allows to include several model architectures, hyperparameters and evaluation metrics that are specific to the problem, data, and desired performance criteria. Multiple parameter realizations are tested in parallel and a subsample of the original problem is used to measure the performance. Once the optimal architecture is chosen according to the image data, the deep learning pipeline performs deconvolution and denoising, focusing and classification tasks \cite{DelliVeneri2023}. 
    
The Bayesian optimization algorithm is supported by a parameter search employing surrogate models: the best set of parameters for a given objective is efficiently found while reducing the number of expensive model evaluations. The expensive objective function (deep learning model) is approximated by its surrogate model, described by Gaussian Processes (GPs): $y=GP(\mu,\Sigma)$ with $\mu(x)$ and $\Sigma(x,x')$ being mean and covariance functions, respectively. The Mat\'ern kernel is chosen as covariance function to capture smoothness and correlations in the data: $\Sigma=K(x,x')=\sigma^2(1+\frac{(x-x')^2}{2\alpha l})^{-\alpha}$ with $\alpha$ and $l$ representing the smoothness and the lengthscale parameters, respectively. The surrogate model is trained on a limited number of initial evaluations of the objective function. In order to target promising regions in the parameter space, an acquisition function is used to decide which set of parameters are evaluated next. In this development the Expected Improvement (EI) has been used as acquisition function: $EI(x) = E [ {\rm max}(f(x) - f(x'), 0) ] $ with $f(x)$ and $f(x')$ indicating the predicted mean of the objective function at point $x$, based on the GP, and the best value observed so far in the optimization process, respectively. EI quantifies the potential gain in performance. It prefers models for which $\mu > f(x')$ (exploitation) or where the standard deviation $\sigma(x)$ is high (exploration). As already shown in \cite{preuss2018}, the key advantage of using surrogate models in Bayesian parameter search is that it reduces the number of costly evaluations of the objective function. 


\subsubsection{Benchmark on archived ALMA data cubes}
\begin{figure}
\vspace{-2cm}
\centering
\rotatebox{-90}{\includegraphics[width=9.cm]{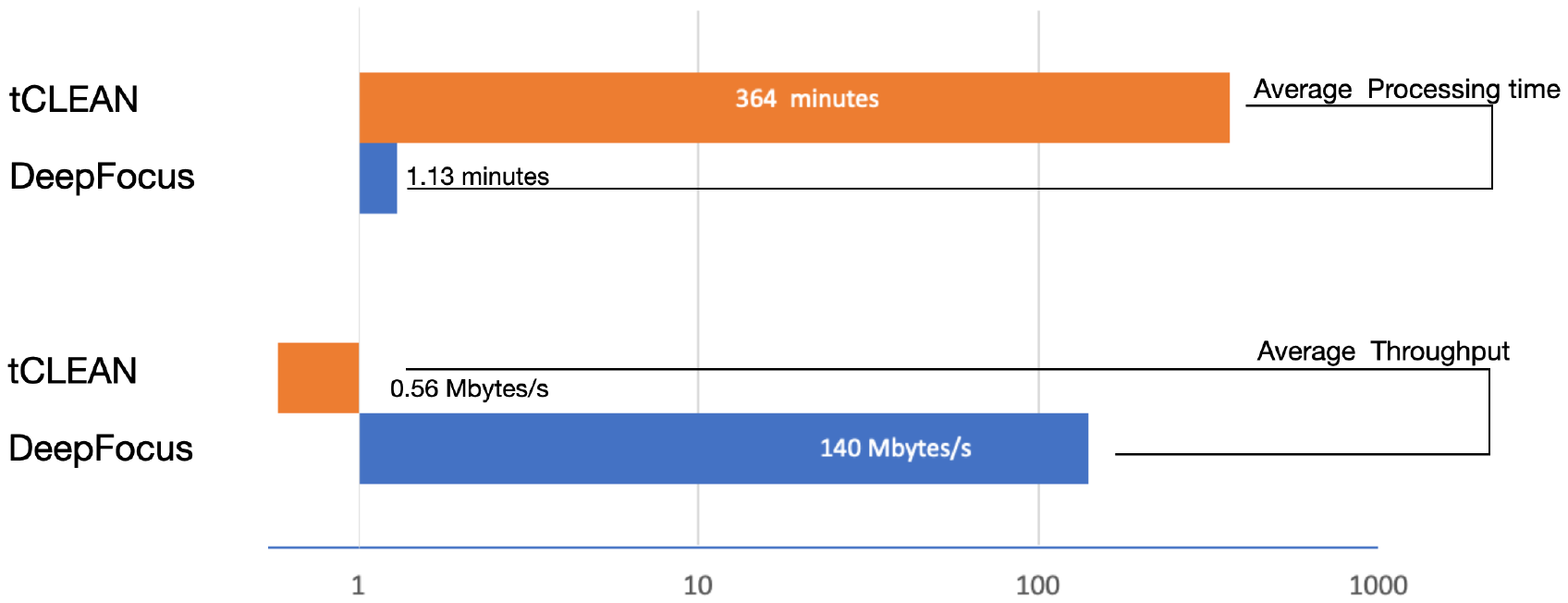}}
\vspace{-2cm}
\caption{Comparison of processing time and computing throughput with tCLEAN and DeepFocus on $29\times10^3$ archived cube data from Cycle 7, 8 and 9. This represents a rough estimate because at this stage of development, it is challenging to make a robust comparison between the two techniques.}
\label{comparough}
\end{figure}

The European ALMA Regional Centre (EU ARC) cluster is frequently utilized for processing the standard pipeline and generating official products delivered to principal investigators.
Using the EU ARC cluster, Deep Focus is executed on $29\times 10^3$ archived ALMA cubes taken during the three most recent Cycles (7--9) (Fig.~\ref{comparough}). We observe an average processing time of $1.13$ minutes per cube and an average compute throughput of $140$ MB/s. 
The same data have been cleaned by tCLEAN algorithm \cite{hogbom,CASA}, see also \cite{Guglielmetti2019, Tychoniec2022}. tCLEAN is executed with the parameter $\text{niter}=1000$, corresponding to the number of cleaning iterations, employing parallel computing. The average computing throughput with tCLEAN resulted of $0.56$ MB/s showing $250$ times lower rate of performance calculations with respect the ML algorithm DeepFocus. The image processing speed improvement achieved by DeepFocus varies significantly depending on the image size, ranging from a 280-fold to a remarkable 5500-fold increase in speed. Employing algorithms as DeepFocus, the image deconvolution process may take only few minutes on large cubes. GPUs, commonly used by ML algorithms, are providing a large benefit for synthesis image analysis. DeepFocus demonstrated high image fidelity and high-performance computing for image reconstruction on ALMA data cubes. 
%
%
%
%
\subsection{A refined ALMA simulator: ALMASim}
In \cite{DelliVeneri2023,Guglielmetti2022} it has been shown that DeepFocus is applied to ALMA dirty cubes, learning from the input data the celestial sources, the noise, the instrumental point spread function (beam). The algorithm allows for extreme data compression by leveraging both spatial and frequency information. DeepFocus being a supervised ML algorithm, the process to build and evaluate the models is essential. To support the needed workflow, the ALMASim package has been developed. 

ALMASim is an extension of the CASA Simulator package \cite{CASA}. We increase the capabilities of this simulator to be tailored for the development of ML imaging algorithms and for quality and reliability assessments. 
The novel package builds upon the CASA PiP Wheels, the MARTINI Package \cite{Martini}, and the Illustris Python Package \cite{Illustris} to facilitate the creation of observations involving high-redshift point-like sources as well as nearby extended sources across the entire spectrum of ALMA configurations. 

\begin{figure}
\centering
\rotatebox{-90}{\includegraphics[width=9.cm]{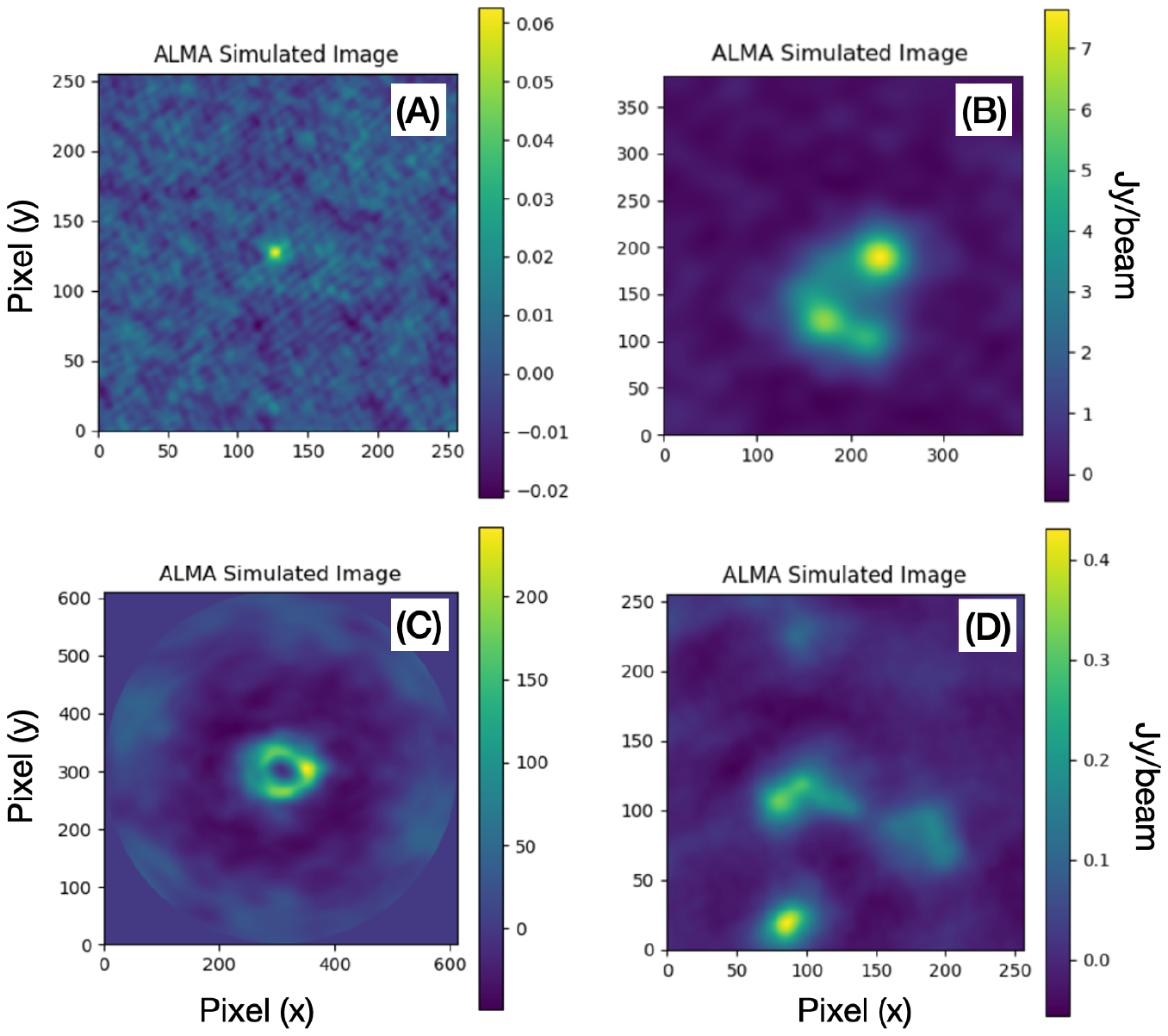}}
\caption{Example of ALMA simulated sources (dirty images) created with the ALMASim package: (A) point like, (B) Gaussian shape, (C) extended, (D) diffuse emissions.}
\label{exasim}
\end{figure}

Following the CASA Simulator, the observational input parameters (precipitable water vapour, band, antenna configuration, bandwidth, integration time, scan time) are provided with the source properties (signal to noise ratio, peak brightness). ALMASim introduces several optional models for the science target brightness distribution: diffuse, point like, Gaussian, extended (Fig.~\ref{exasim}). For the diffuse emission the Numerical Information Field Theory package (NIFTy) \cite{Nifty} is used. The simulated science target distributions enter into the MARTINI package \cite{Martini} to transform the input data to radio observations. Mock observations of simulated celestial sources (as high redshift point sources, low redshift galaxies, extended structures, both continuum and emission lines in ALMA cubes) are produced efficiently. 
In output, the novel package generates Sky model cubes and Dirty ALMA cubes, both in FITS format. The Sky models represent the simulated real sky without any noise or instrumental artefacts added to the image. The dirty ALMA cubes correspond to the Fourier inversion of the observed visibilities corrupted by noise (e.g. instrumental, thermal). ALMASim can optionally provide output such as the ALMA dirty beam, Measurement Sets files, and plots of 2D integrated cubes and 1D spectra for all simulated data.

We aim at having this service available to the scientific community as open source. A number of pre-configured parameter sets will be available to generate identical datasets on which to test algorithms. 
ALMASim empowers users in generating synthetic datasets designed for the development of deconvolution and source detection models. Each user will be able to generate the dataset on its local machine. The package is engineered to harness the usage of MPI parallel computing, particularly on contemporary HPC clusters, to efficiently produce thousands of ALMA data cubes and their visibilities. Users have the choice to define the input parameters or to opt for a completely randomized configuration and sources. 

\subsubsection{Sharpening mock data with realistic noise characteristics}
We want to ensure that the simulated data closely resemble actual ALMA observations. The generation of synthetic noise has to reproduce the noise in real ALMA observations and introduce those characteristics in the artificial images. 
Two approaches are used: (1) adding noise to the generated ALMA measurement sets, (2) empirical noise modelling. 
Realistic noise characteristics in artificial ALMA images is commonly added to the generated ALMA measurement sets, which contain the visibility data and calibration information from the simulated observations. Synthetic noise can be generated from the CASA Simulator capable to corrupt the visibility data with thermal noise and atmospheric attenuation, and with leakages (cross polarization). Corruption with atmospheric turbulence, or adding gain fluctuations or drift, are also possible within the CASA Simulator. However, correlated noise (e.g., due to antenna placement) is not straightforwardly accounted, but possible while corrupting the measurements sets. Once the several components are introduced, the final noisy image is verified such that the noise levels in the synthetic image match the noise characteristics of the desired simulated observations. Alternatively, the empirical approach to noise modeling is not bounded to any theoretical model, allowing the approach to be applicable at other observatories and wavelengths. 

\begin{figure}
\vspace{-3cm}
\begin{adjustwidth}{-\extralength}{0cm}
\centering
\rotatebox{0}{\includegraphics[width=12cm]{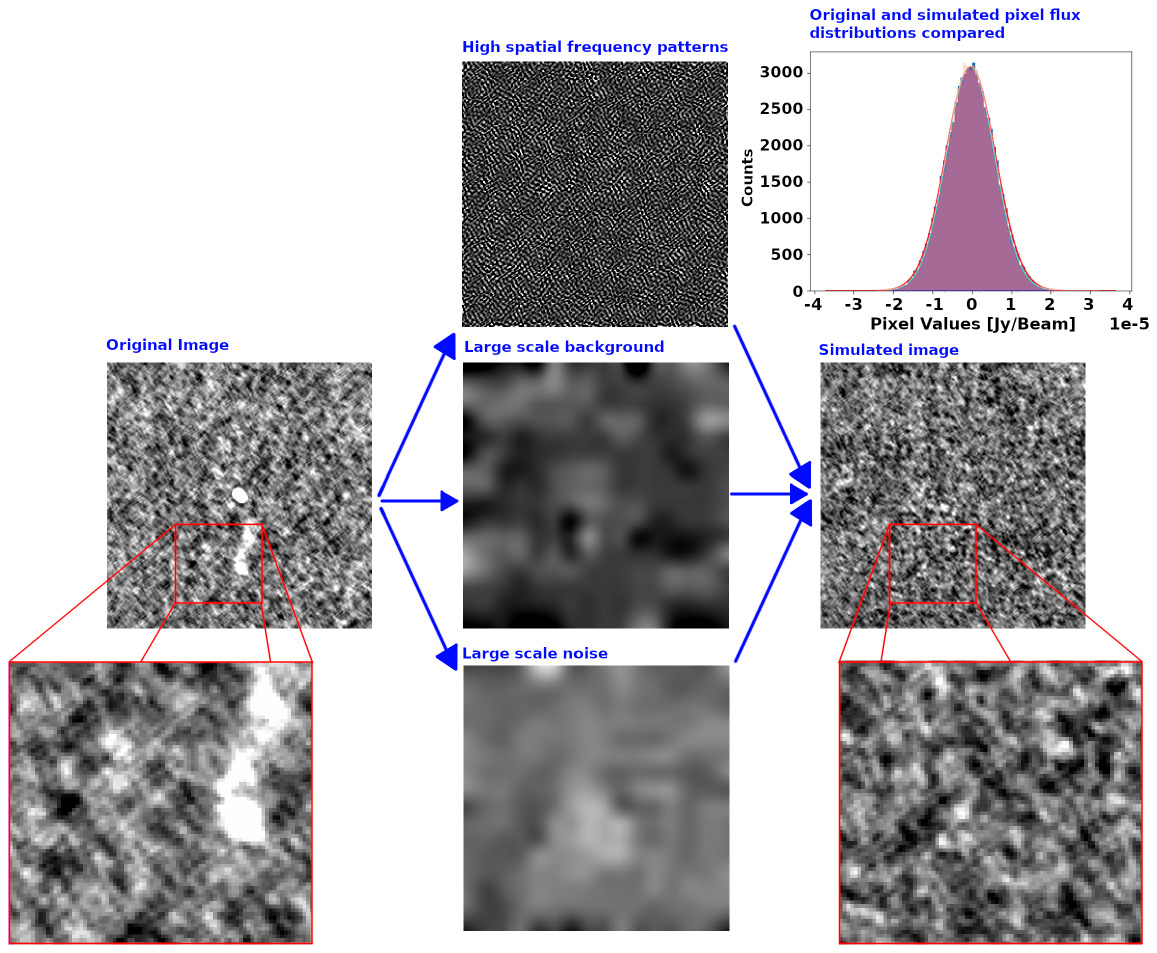}}
\end{adjustwidth}
\vspace{-3cm}
\caption{Simplified visual explanation of the empirical approach to the noise modelling: different background and noise components, measured at scales larger and shorter than the typical beam scale (central panels), are isolated from a real ALMA image (e.g., an ALMA calibrator, in the left panel) and then added to a simulated image (right panel). In this example, we considered local fluctuations (center, bottom panel), large scale background (central panel), and  high spatial frequency patterns (center, top panel). Instead of using a theoretical model to simulate noise and instrumental response, the same effects are directly measured from real observations obtained in comparable situations (telescope configuration and atmospheric conditions).
}
\label{noise}
\end{figure}

\paragraph{Empirical Noise Modeling}
Generation of synthetic noise is limited in reproducing all complexities encountered in real image data, though we want the simulated ALMA images to closely resemble actual ALMA observations. Commonly, quality assurance for principal investigator's parameter of interest (as sensitivity, resolution) occurs on image space. 
We investigate the possibility to learn the noise patterns from real ALMA data with the goal to understand the noise properties, such as the spatial and spectral distribution of noise, and possibly consider correlated noise. Therefore, rather than relying solely on artificial noise added to a simulated real Sky, an empirical approach to noise modelling is used and applied to ALMASim to encompass a broader range of noise components.
Noise components are characterized at larger scales and at scales similar to and smaller than the ALMA beam size from observed data. By extracting high spatial-frequency noise patterns present in the image, we try to capture 
the complexity of the noise environment.
When replicated to create the new simulated image, we take care to preserve the statistical properties of the flux distribution on individual pixels, matching those observed in the original image (except in regions originally occupied by 
the astronomical sources of interest). In Fig.~\ref{noise} a simulated ALMA image characterized by noise is shown and extracted from a real ALMA image. The real ALMA image displays a quasi stellar object with a jet like structure. The empirical approach assists in reinforcing the creation of realistic ALMA simulations. Extracting noise from ALMA observations and integrating it into simulated Sky data improves the fidelity of noise modeling, leading to a more precise representation of the intricate noise characteristics often encountered in ALMA observations.  

\section{Outlook and Conclusions}
%
The ALMA2030 roadmap is bringing a significant transformation due to the increment of data rates and volume. 
The ESO internal ALMA development study BRAIN provided an exploration of concepts for advanced imaging techniques. RESOLVE and DeepFocus are applicable to large data volumes while requiring the least amount of human intervention. \\
While RESOLVE stands out for its exceptional capacity to detect extended emissions and distinguish them from point sources, it will naturally outclass at combining data from various measurement sets for group-level imaging. Group-level imaging allows us to combine data from multiple arrays (12-m, 7-m and total power) to capture both high-resolution and low-resolution features in a single observation. RESOLVE appears to be emerging as the preferred tool in the scientific community. Currently, the RESOLVE algorithm is in development to enhance its efficiency in processing ALMA cube images.
On the other hand, DeepFocus is well-equipped with essential features to be used in operations: e.g.~real time processing, cube imaging of all spectral windows for large datasets. DeepFocus is actively advancing in its capability to detect diverse source morphologies and it is planned to learn from the freshly archived data. Its companion package, ALMASim, opens doors to the creation of tailored imaging algorithms, specifically designed to address unique scientific cases. Although we still not have a clear design of ALMA2030, DeepFocus is foreseeable to perform image processing soon after the calibrated data are reduced in real time processing. If freshly observed data are calibrated and ingested in the archive, the DeepFocus algorithm could efficiently perform the imaging of the calibrated data within few minutes. A real time imaging service from a science platform of an ALMA archive is realistic. During the ALMA imaging process, if structures are present in the images caused by, e.g., a glitch, then an alert system could be sent to the system astronomers to perform the needed investigations and take actions for the following ALMA observations and/or an alert system can be sent to the next generation CASA team \cite{CNGI}. If the archive is the central system to acquire the ingested calibrated data, imaging algorithms of choice can be employed to perform the imaging of the measurement set and at the group level when available for each project.

Enhancing the efficiency of the observatory and establishing a comprehensive archive in terms of products will greatly enrich data mining opportunities for the scientific community.




\authorcontributions{Conceptualization, F.G.; methodology, F.G., E.V., T.E., G.L., M.D.V., I.B., L.T., F.S., J.R.; software, M.D.V., V.J., C.B., A.D., J.R., ; validation, F.S., L.T.; formal analysis, M.D.V., V.J.; resources, F.G., T.E., G.L., M.D.V.; writing---original draft preparation, F.G.; writing---review and editing, T.E., F.S., E.V., M.D.V. All authors have read and agreed to the published version of the manuscript.}

\dataavailability{Publicly available datasets were analyzed in this study. The data can be found querying the ALMA Archive: \url{https://almascience.eso.org/aq/}}

\acknowledgments{This research is supported by an ESO internal ALMA development study investigating interferometric image reconstruction methods.The authors would like to acknowledge the Max Planck Computing and Data Facility (MPCDF) for providing computational resources and support for this research. J.R.~acknowledges financial support by the German Federal Ministry of Education and Research (BMBF) under grant 05A20W01
(Verbundprojekt D-MeerKAT).}

\conflictsofinterest{The authors declare no conflict of interest.} 

\abbreviations{Abbreviations}{
The following abbreviations are used in this manuscript:\\

\noindent 
\begin{tabular}{@{}ll}
ALMA & Atacama Large Millimeter/submillimeter Array\\
ALMA2030 & ALMA Development Roadmap\\
BRAIN & Bayesian Reconstruction through Adaptive Image Notion\\
DSHARP & Disk Substructures at High Angular Resolution Project \\
EI & Expected Improvement \\
ESO & European Southern Observatory\\
EU ARC & European ALMA Regional Centre \\
IF & Intermediate Frequency bandwidth \\
FITS & Flexible Image Transport System \\
GP & Gaussian Process \\
GPU & Graphics Processing Unit \\
HPC & High-Performance Computing \\
ML & Machine Learning 
\end{tabular}
\newpage
\begin{tabular}{@{}ll}
MPI & Message Passing Interface \\
NIFTy & Numerical Information Field Theory \\
pdf & probability density function \\
WSU & Wideband Sensitivity Upgrade
\end{tabular}
}

\begin{adjustwidth}{-\extralength}{0cm}

\reftitle{References}

\PublishersNote{}
\end{adjustwidth}
\end{document}